\documentclass[a4paper,11pt]{article}

\usepackage{jheppub} 

\usepackage[T1]{fontenc} 
\usepackage[english]{babel}
\usepackage[utf8]{inputenc}
\usepackage{amsmath,amsthm}

\usepackage{etoolbox}

\title{\boldmath Yang-Baxter Deformations of the $AdS_{5}\times S^{5}$ Pure Spinor Superstring }

\author[]{H\'ector A. Ben\'itez,}
\author[]{Victor O. Rivelles}

\affiliation[]{ Instituto de F\'isica, Universidade de S\~ao Paulo\\Rua do Mat\~ao, 1371, 05508-090, S\~ao Paulo, SP, Brazil}

\emailAdd{habenite@if.usp.br}
\emailAdd{rivelles@fma.if.usp.br}

\abstract{We present integrable Yang-Baxter deformations of the $AdS_{5}\times S^{5}$ pure spinor superstring theory which were obtained by using homological perturbation theory. 
Its equations of motion and BRST symmetry are discussed and its Lax connection is derived. We also show that its  target space background is the same generalized supergravity background found for Yang-Baxter deformations of the Green-Schwarz superstring in $AdS_{5}\times S^{5}$. }

\theoremstyle{definition}

\theoremstyle{plain}
\newtheorem{theorem}{Theorem}

\newtheorem{prop}{Proposition}

\begin{document} 

\makeatletter 
\patchcmd{\maketitle}{\@fpheader}{}{\hfill}{} 
\makeatother 

\maketitle
\flushbottom

\section{Introduction}

There are many instances in the context of the AdS/CFT correspondence where both sides of the duality present an integrable structure \cite{Beisert,Bombardelli}. On the string theory side it is well known that the $AdS_{5}\times S^{5}$ superstring equations of motion either in the Green-Schwarz (GS) or in the pure spinor (PS) formulation can be cast into a zero curvature equation satisfied by a Lax pair 
\cite{Polchinski,Vallilo}. 
More recently, significant progress has been made in deforming the $AdS_{5}\times S^{5}$ structure while preserving its integrability and the fermionic $\kappa$-symmetry. On one hand, the GS $\lambda$-deformed models \cite{Hollowood} are based on a $G/G$ gauged WZW model and yields a target superspace corresponding to a supergravity background. On the other hand, the GS Yang-Baxter (YB) deformed models make use of a linear operator, called $R$-matrix, which solves the modified classical Yang-Baxter equation (mCYBE) (also known as $\eta$-deformations) \cite{vicedo,delduc}, or solves the homogeneous CYBE \cite{kawa,matsu}. 
Even though $\kappa$-symmetry is preserved for these deformations, its target superspace does not solve the equations of motion of type IIB supergravity \cite{smatrix,puzzles,Hoare:2016hwh} but rather a generalization of them.
It was also argued that, even without a dilaton to preserve Weyl invariance at one-loop level, the generalized supergravity backgrounds still define a two-dimensional scale invariant theory \cite{scale}. This apparent conflict was solved by Tseytlin and Wulff \cite{kappa} who showed that contrary to the standard assumption that $\kappa$-symmetry implies the equations of motion of type IIB supergravity, it in fact leads to generalized supergravity equations of motion.
In this framework, the supergravity equations of motion can be recovered 
when the $R$-matrix 
is unimodular \cite{target}. 

These generalized backgrounds are also related to the standard supergravity equations by $T$-dualizing a supergravity target space in a isometric direction which is a symmetry of all the fields except for the dilaton  transforming linearly in this direction \cite{Hoare:2015wia,Hoare1}. Also, the relation between homogeneous YB deformations and T-duality have been extensively studied \cite{vanTongeren,vanTongeren:2015uha,Osten,Hoare2,Hoare3}. It has also been shown that these deformations and the T-duality transformations of the original model can be formulated in a unified description \cite{Borsato:2016,Borsato:2017,Borsato:2018}. The emergence of generalized supergravity backgrounds has also been further explored in the context of open-closed string map and in the double field theory formalism \cite{vanTongeren:2015uha,Araujo1,Bakhmatov:2017joy,Araujo:2018rbc,Sakamoto:2017wor,Fernandez-Melgarejo:2017oyu,Lust,Sakamoto,vanTongeren:2016eeb}.

In the PS formalism the world-sheet metric is in the conformal gauge and the $\kappa$-symmetry of the GS superstring is replaced by a global BRST symmetry avoiding well known issues with the light-cone gauge. This provides a powerful tool to gain new insights like properties of  vertex operators and correlation functions  \cite{Mikhailov1, Berkovits5, Ramirez}. It also allowed to show that the integrability of the PS string in  $AdS_{5}\times S^{5}$ \cite{Vallilo} persists to all loops in the quantum theory \cite{Berkovits4}. Regarding deformations of PS superstrings 
%
just a few cases are known, like 
the $\beta$-deformation, obtained by a TsT  transformation on the supergravity background \cite{beta}, 
or the $\lambda$-deformation  for the matter sector of the pure spinor model \cite{david}. 

A more systematic way to deform the PS superstring in $AdS_{5}\times S^{5}$ was proposed in \cite{Bevilaqua} and makes use of homological perturbation theory. 
The main idea is to find a deformed action and a deformed nilpotent BRST operator which can be constructed as a series expansion in the deformation parameter. At first order it was found that the action is proportional to an integrated vertex operator parametrized by an antisymmetric $R$-matrix. When acting on the matter sector, the deformed BRST operator has the same structure as the $\eta$-deformed $\kappa$-symmetry of the GS superstring \cite{target}, thus suggesting that the PS deformed model could give rise to an $\eta$-deformed background. Moreover, nilpotency of the BRST charge at first order implies the CYBE or the mCYBE for the $R$-matrix \cite{Bevilaqua}.
It was also found that in  the flat space limit BRST invariance is not enough to characterize the linearised equations of motion for type IIB supergravity \cite{cornering}. This is due to a  conflict with the conformal invariance of the deformed PS action. In order to preserve the conformal symmetry for the deformed world-sheet theory the corresponding vertex operator must be given by a primary operator of $AdS_{5}\times S^{5}$. This means that the double pole of the OPE for the vertex operator and the energy-momentum tensor, which is proportional to the action of the Laplacian on $\mathfrak{psu}(2,2|4)$, must vanish \cite{vertices}. As shown in \cite{Bevilaqua} this requirement implies the unimodular condition for the $R$-matrix. It has also been pointed out that the cohomology of the deformed BRST charge in the flat space limit  does not give rise to a supergravity background but rather to a generalized supergravity one \cite{ghost,araujo}. Since these results were found at first order in the deformation parameter it would be very interesting to see which of these properties hold for the fully deformed theory. Besides that, since the YB deformations of the GS superstring in $AdS_{5}\times S^{5}$ do not require a small deformation parameter we would expect the same to be true for the PS case. 

The aim of this paper is to extend the approach of \cite{Bevilaqua} to all orders in the deformation parameter to obtain full YB deformations of the PS superstring in $AdS_{5}\times S^{5}$. 
For simplicity we will consider only the case in which the $R$-matrix satisfies the CYBE. The case where the  
$R$-matrices satisfy the mCYBE should follow the same lines as in \cite{Bevilaqua}. A troublesome issue of this construction is that the BRST charge acts through a non-local operator on the anti-field sector turning the full deformed action non-polynomial. We then find that it is possible to remove the anti-field sector so that the complete deformed action is a polynomial expansion in the $R$-matrix.  As a consequence, the BRST operator has now an infinite series expansion in the ghost sector and its nilpotency holds only on-shell.  
Even so, this local action allows us to read the background superfields using the general action of Berkovits and Howe \cite{BH}. By rewriting the PS deformed action in terms of the GS deformed variables we show that  both, the GS and the PS YB deformed models have the same geometry and target space fields. 

The YB deformations of the GS superstring can be constructed in a way which manifestly preserves the integrability of the undeformed theory \cite{delduc}. 
As expected \cite{vicedo,delduc,vanTongeren:2018vpb}, it is possible to recast the \textit{deformed} equations of motion of the PS superstring 
in such a way that it presents the same algebraic structure as the \textit{undeformed} ones. This allow us to find a Lax  representation for them establishing integrability for the PS case.


The contents of this paper is the following. In section 2 we review some important properties about the PS superstring in $AdS_{5}\times S^{5}$ like its BRST symmetry and the Lax representation for the equations of motion. In section 3 we present the full deformation of the $AdS_{5}\times S^{5}$ PS superstring extending the linearised results of  \cite{Bevilaqua} to all orders. Then, in section 4, we show how to get rid of the awkward non-polynomial terms appearing in the deformed action. In section 5 we show the integrability of the deformed action by constructing a suitable Lax connection. Finally, in section 6, we make contact with the GS deformed superstring and show that the target space superfields of our model correspond to those of generalized supergravity.

\section{Review of the pure spinor superstring in $AdS_{5}\times S^{5}$ }

It is well known that the superstring theory in $AdS_{5}\times S^{5}$ can be formulated  as a supercoset sigma model on $PSU(2,2|4)/(SO(4,1)\times SO(5))$ \cite{Mazzucato}. In this construction, a key  role  is played  by  the $\mathcal{Z}_{4}$ grading of the superalgebra $\mathfrak{psu} (2,2|4)$ which allows us to decompose it as  
\begin{equation}
\mathfrak{g} = \mathfrak{g}_{0}+\mathfrak{g}_{1}+\mathfrak{g}_{2}+\mathfrak{g}_{3}\,.
\label{}
\end{equation} 
If $g$ is an element of the supergroup $PSU(2,2|4)$ we define the Maurer-Cartan form
as $A=-dg\,g^{-1}$. Since it takes values in $\mathfrak{psu}(2,2|4)$  we can decompose it as 
\begin{equation}
A = -dg\,g^{-1} = A_{0}+A_{1}+A_{2}+A_{3}\,.
\label{aja}
\end{equation}

The pure spinor action in $AdS_{5}\times S^{5}$ is given by \cite{Berkovits6} \footnote{In our notation a $\mathfrak{psu}(2,2|4)$-valued field $X_{i\,+}$ ($X_{i\,-}$) takes values in $\mathfrak{g}_{i}$ and has conformal dimension (1,0) ((0,1)). We name the generators according to their grading $\mathfrak{g}_{0}=\{t_{ab} \}$, $\mathfrak{g}_{2}=\{t_{a} \}$, $\mathfrak{g}_{1}=\{t_{\alpha} \}$, $\mathfrak{g}_{3}=\{t_{\hat{\alpha}} \}$. An exhaustive discussion of the pure spinor superstring action can be found in \cite{Mazzucato} and references therein.}  
\begin{equation}
S_{AdS}=\int Str \left(\frac{1}{4}A_{+}d_{PS}A_{-}+\omega_{1+}\partial_{-}\lambda_{3}+\omega_{3-}\partial_{+}\lambda_{1}+N_{0+}A_{0-}+N_{0-}A_{0+}-N_{0-}N_{0+}\right)\,,
\end{equation}
with $d_{PS}=P_{1}+2P_{2}+3P_{3}$, where $P_{i}$ projects an element of the superalgebra $\mathfrak{g}$ on its $\mathfrak{g}_{i}$-component. The Lie algebra valued ghost fields are defined as
\begin{equation}
\lambda_{1}=\lambda^{\alpha}t_{\alpha}\,,\quad \omega_{3-}=\omega_{\alpha}\eta^{\alpha\hat{\alpha}}t_{\hat{\alpha}}\,,\quad \lambda_{3}=\hat{\lambda}^{\hat{\alpha}}t_{\hat{\alpha}}\,,\quad \omega_{1+}=\hat{\omega}_{\hat{\alpha}}\eta^{\alpha\hat{\alpha}}t_{\alpha}\, ,
\end{equation}
where $\eta^{\alpha\hat{\alpha}}=-\eta^{\hat{\alpha}\alpha}$ is numerically equal to the identity matrix. 
The bosonic ghosts $\lambda^{\alpha}$ and $\hat{\lambda}^{\hat{\alpha}}$ are constrained to satisfy the pure spinor condition
\begin{equation}
\lambda\gamma^{a}\lambda=\hat{\lambda}\gamma^{a}\hat{\lambda}=0\,,
\label{purespinor}
\end{equation}
and the pure spinor Lorentz currents are given by  
\begin{equation}
N_{0-}=-\{\omega_{1+},\lambda_{3}\}\,, \qquad N_{0+}=-\{\omega_{3-},\lambda_{1}\}\,.
\end{equation}

The action is invariant under a BRST symmetry whose classical charge $Q=Q_{L}+Q_{R}$ is given by
\begin{equation}
Q_{L}=\oint \mbox{Str}(\lambda_{1}A_{3-})\,,\qquad Q_{R}=\oint \mbox{Str}( \lambda_{3}A_{1+})\,,
\label{BRSTads}
\end{equation}
and acts on a group element as a derivative
\begin{eqnarray}
\epsilon\,Q(g)&=&(\epsilon\lambda_{1}+\epsilon\lambda_{3})g\,,\label{brs2.1}\\
\epsilon\,Q(w_{3-})&=&-\epsilon A_{3-}\,,\label{bras1}\\
\epsilon\,Q(w_{1+})&=&-\epsilon A_{1+}\,,
\label{brs2}
\end{eqnarray}
implying 
\begin{eqnarray}
\epsilon\,Q(A_{i-})&=&\delta_{i,1}\partial_{-}(\epsilon\lambda_{1})+[A_{i+3;-},\epsilon\lambda_{1}]+\delta_{i,3}(\partial_{-}\epsilon\lambda_{3})+[A_{i+1;-},\epsilon\lambda_{3}]\,,\\
\epsilon\,Q(A_{i+})&=&\delta_{i,1}\partial_{+}(\epsilon\lambda_{1})+[A_{i+3;+},\epsilon\lambda_{1}]+\delta_{i,3}(\partial_{+}\epsilon\lambda_{3})+[A_{i+1;+},\epsilon\lambda_{3}]\,,\\
\epsilon\,Q(N_{0-})&=&[A_{3-},\epsilon\lambda_{1}]\,,\\
\epsilon\,Q(N_{0+})&=&[A_{1+},\epsilon\lambda_{3}]\,.
\end{eqnarray}

For the matter sector the equations of motion are obtained from variations $\delta g=g\xi_{i}$, $i=1,2,3$, where $\xi_{i}$ is an element of $\mathfrak{g}_{i}$. Defining the covariant derivatives as
\begin{equation}
D_{\pm}=\partial_{\pm}+[A_{0\pm},\,\,]\,, 
\end{equation}
we find that
\begin{eqnarray}
D_{-}A_{1+}&+&[A_{1-},N_{0+}]-[N_{0-},A_{1+}]=0\,,\label{desde}\\
D_{-}A_{2+}&+&[A_{1-},A_{1+}]+[A_{2-},N_{0+}]-[N_{0-},A_{2+}]=0\,,\\
D_{-}A_{3+}&+&[A_{1-},A_{2+}]+[A_{2-},A_{1+}]-[A_{3-},N_{0+}]-[N_{0-},A_{3+}]=0\,,\\
D_{+}A_{1-}&+&[A_{2+},A_{3-}]+[A_{3+},A_{2-}]+
[A_{1-},N_{0+}]-[N_{0-},A_{1+}]=0\,,\\
D_{+}A_{2-}&+&[A_{3+},A_{3-}]+[A_{2-},N_{0+}]-[N_{0-},A_{2+}]=0\,,\\
D_{+}A_{3-}&+&[A_{3-},N_{0+}]-[N_{0-},A_{3+}]=0\,.
\label{hasta}
\end{eqnarray}
Similarly, the equations of motion for the ghost sector are obtained by varying the action with respect of $\lambda$ and $\omega$ and expressing the result in terms of the Lorentz currents 
\begin{equation}
D_{\pm}N_{0\mp}-[N_{0\pm},N_{0\mp}]=0\,.
\label{Nec}
\end{equation}

Classical integrability can be proven by constructing the Lax pair \cite{Vallilo}
\begin{eqnarray}
L_{+}(z)&=&A_{0+}+z^{-3}A_{1+}+z^{-2}A_{2+}+z^{-1}A_{3+}+(z^{-4}-1)N_{0+}\,,\nonumber\\
L_{-}(z)&=&A_{0-}+z A_{1-}+z^{2}A_{2-}+z^{3}A_{3-}+(z^{4}-1)N_{0-}\,,
\label{laxundef}
\end{eqnarray}
where $z$ is the spectral parameter, in such a way that the equations of motion \eqref{desde}-\eqref{hasta} and \eqref{Nec} are equivalent to the zero curvature condition
\begin{equation}
\partial_{-}L_{+}-\partial_{+}L_{-}+[L_{-},L_{+}]=0\,.
\end{equation}
Defining $z=e^{l}$ it is possible to express the density of the  conserved currents as
\begin{equation}
j=g^{-1}\left(\frac{dL}{dl}\Big|_{l=0}\right)g\,,
\end{equation}
such that $\partial_{+}j_{-}-\partial_{-}j_{+}=0$. Explicitly, the $j_{\pm}$ currents are
\begin{eqnarray}
j_{-}&=& g^{-1}(A_{1-}+2A_{2-}+3A_{3-}+4N_{0-})g\,,\nonumber\\
j_{+}&=& -g^{-1}(3A_{1+}+2A_{2+}+A_{3+}+4N_{0+})g\,.
\label{current}
\end{eqnarray}

It is also useful to find the BRST transformation  of the currents
\begin{eqnarray}
\epsilon Q(j_{+})&=&\partial_{+}\Lambda(\epsilon)+4g^{-1}(D_{+}\epsilon\lambda_{1}-[N_{0+},\epsilon\lambda_{1}])g\,,\nonumber\\
\epsilon Q(j_{-})&=&\partial_{-}\Lambda(\epsilon)-4g^{-1}(D_{-}\epsilon\lambda_{3}-[N_{0-},\epsilon\lambda_{3}])g\,,
\label{Qj}
\end{eqnarray}
where,
\begin{equation}
\Lambda(\epsilon)= g^{-1}(\epsilon\lambda_{1}-\epsilon\lambda_{3})g\,.
\end{equation}

We recall that the BRST charge is nilpotent up to equations of motion \cite{Berkovits4}. In order to  make the  BRST charge nilpotent off-shell we introduce a pair of fermionic anti-fields $(\omega_{1+}^{*}, \omega_{3-}^{*})$ which must satisfy the following condition \cite{Vafa}
\begin{equation}
\{\lambda_{1},\omega_{1+}^{*}\}=\{\lambda_{3},\omega_{3-}^{*}\}=0\,.
\label{vincanti}
\end{equation}
The BRST transformations \eqref{bras1} and \eqref{brs2} are then be modified to
\begin{equation}
\epsilon Q(w_{3-})=-\epsilon A_{3-}-\omega_{3-}^{*}\,,\quad \epsilon Q(w_{1+})=-\epsilon A_{1+}-\omega_{1+}^{*}\,,\label{ppp}
\end{equation}
and the BRST transformation of the anti-fields is given by
\begin{equation}
\epsilon Q(\omega_{3-}^{*})=D_{-}\epsilon\lambda_{3}-[N_{0-},\epsilon\lambda_{3}]\,,\quad \epsilon Q(\omega_{1+}^{*})=D_{+}\epsilon\lambda_{1}-[N_{0+},\epsilon\lambda_{1}]\,.
\label{pp}
\end{equation}
Notice that the BRST transformations of the conserved currents $j_{\pm}$ are proportional to the equations of motion for $\lambda$, therefore the conserved charge is BRST invariant only when the classical equations of motion hold. It is possible to avoid this situation by including the anti-fields in the  currents \eqref{current} 
\begin{eqnarray}
\tilde{j}_{-}&=& g^{-1}(A_{1-}+2A_{2-}+3A_{3-}+4N_{0-}+4\omega_{3-}^{*})g\,,\label{j+}\\
\tilde{j}_{+}&=& -g^{-1}(3A_{1+}+2A_{2+}+A_{3+}+4N_{0+}+4\omega_{1+}^{*})g\,.\label{j-}
\end{eqnarray}
which transform under \eqref{brs2.1}, \eqref{ppp} and \eqref{pp} as
\begin{equation}
\epsilon Q(\tilde{j}_{\pm})=\partial_{\pm}\Lambda(\epsilon)\,.
\label{Qbarj}
\end{equation}
 We then find that the action 
\begin{equation}
S_{0}=\int\left(\frac{1}{4}A_{+}d_{PS}A_{-}+\omega_{1+}\partial_{-}\lambda_{3}+\omega_{3-}\partial_{+}\lambda_{1}+N_{0+}J_{0-}+N_{0-}J_{0+}-N_{0-}N_{0+}-\omega_{1+}^{*}\omega_{3-}^{*}\right), 
\label{so}
\end{equation}
is invariant under the BRST transformations \eqref{brs2.1}, \eqref{ppp} and \eqref{pp}.

\section{Deformation of the  $AdS_{5}\times S^{5}$ pure spinor superstring}

In this section we will follow the general deformation procedure for a BRST invariant action presented in \cite{Bevilaqua}. 
The deformed action and BSRT charge are constructed as a series expansion in $\eta$ 
 \begin{eqnarray}
 S_{def}&=&S_{0}+\eta\int V^{}_{1}+\eta^{2}\int V^{}_{2}+\ldots\,,\\
 Q_{def}&=&Q_{0}+\eta Q_{1}+\eta^{2}Q_{2}+\ldots\,,
 \end{eqnarray}
where the coefficients of the expansion are determined by imposing BRST invariance
 \begin{equation}
 Q_{def}S_{def}=0\,.
 \end{equation}
We find, up to order $\eta^{3}$, that 
 \begin{eqnarray}
 &Q_{0}&S_{0}=0\,,\\
 &Q_{1}&S_{0}+Q_{0}\int V^{}_{1}=0\,,\label{since}\\
 &Q_{2}&S_{0}+Q_{1}\int V^{}_{1}+Q_{0}\int V^{}_{2}=0\,,\label{to}\\
 &Q_{3}&S_{0}+Q_{2}\int V^{}_{1}+Q_{1}\int V^{}_{2}+Q_{0}\int V^{}_{3}=0\,.\label{hastaaa}
 \label{serial}
 \end{eqnarray}
The first of these equations is satisfied since the undeformed model is assumed to be BSRT invariant. In \cite{Bevilaqua} equations \eqref{since} and \eqref{to} were solved and consequently $ V^{}_{1}$, $V^{}_{2}$, $Q_{1}$ and $Q_{2}$ were obtained. Here we review this procedure and solve the equations to all orders. In particular, we will show that the expansion of the BRST charge stops at first order. 

The first step is to make an ansatz to solve \eqref{since} as \cite{Bevilaqua}
\begin{equation}
V^{}_{1}=\frac{1}{4}\int Str(R\tilde{j}_{+},\tilde{j}_{-})\,,
\label{hatjj}
\end{equation}
with $\tilde{j}_{\pm}$ as in \eqref{j+} and \eqref{j-}, and the $R$-matrix being antisymmetric.  
Taking the BRST transformation of $\tilde{j}_{\pm} $ \eqref{Qbarj} we find 
\begin{equation}
Q_{0}V^{}_{1}=-\frac{1}{4}\big(Str(\tilde{j}_{+}, \partial_{-}R\Lambda)-Str(\tilde{j}_{-}, \partial_{+}R\Lambda)\big)\,.
\label{prim}
\end{equation}
To cancel this contribution at first level, $Q_{1}$ is required to satisfy $Q_{1}S_{0}=\frac{1}{4}\big(Str(\tilde{j}_{+}, \partial_{-}R\Lambda)-Str(\tilde{j}_{-}, \partial_{+}R\Lambda)\big)$.
This is achieved by taking $Q_{1}$ as \cite{Bevilaqua}
\begin{eqnarray}
\epsilon Q_{1}(g)&=&gR\Lambda(\epsilon)\,,\\
\epsilon Q_{1}(\omega_{1+}^{*})&=&\mathcal{P}_{13}(g(R\partial_{+}\Lambda(\epsilon)) g^{-1})_{1}\,,\label{antg1}\\ \epsilon Q_{1}(\omega_{3-}^{*})&=&\mathcal{P}_{31}(g(R\partial_{-}\Lambda(\epsilon)) g^{-1})_{3}\,,\label{antg2}
\end{eqnarray}
where the projectors $\mathcal{P}_{13}$ and $\mathcal{P}_{31}$ are defined as
\begin{eqnarray}
\mathcal{P}_{13}A_{1}&=&A_{1}+[\lambda_{3},S_{2}]\,, \\
\mathcal{P}_{31}A_{3}&=&A_{3}+[\lambda_{1},S_{2}]\,, 
\end{eqnarray}
where $S_{2}$ is a vector of grading 2, such that \cite{Mikhailov2}
\begin{equation}
[\lambda_{1},\mathcal{P}_{13}A_{1}]=0\,,\qquad [\lambda_{3},\mathcal{P}_{31}A_{3}]=0\,.
\end{equation}
Some important identities involving the projectors $\mathcal{P}$ can be found in Appendix A.


The next step is to discuss the nilpotency of the BSRT charge. Noticing that $\epsilon Q_{0}g=g\bar{\Lambda}(\epsilon)$ for $\bar{\Lambda}(\epsilon)=g^{-1}(\epsilon\lambda_{1}+\epsilon\lambda_{3})g$ and $\epsilon Q_{1}g=gR\Lambda(\epsilon)$ then  $[\Lambda(\epsilon),\bar{\Lambda}(\epsilon')]=0$ is a consequence of the pure spinor condition. We also find that
\begin{eqnarray}
\epsilon' Q_{1}(\epsilon Q_{0}(g))&=&g\bar{\Lambda}(\epsilon)R\Lambda(\epsilon')\,,\nonumber\\
\epsilon' Q_{0}(\epsilon Q_{1}(g))&=&g\bar{\Lambda}(\epsilon')R\Lambda(\epsilon)+gR[\Lambda(\epsilon),\bar{\Lambda}(\epsilon')]
=-g\bar{\Lambda}(\epsilon)R\Lambda(\epsilon')\,,
\end{eqnarray}
so that $\{Q_{0},Q_{1} \}=0$. We also need to find the action of $Q^{2}_{1}$ over $g$. We start with
\begin{eqnarray}
\epsilon' Q_{1}(\epsilon Q_{1}(g))&=&g\left([R\Lambda(\epsilon'),R\Lambda(\epsilon)]-R\left([R\Lambda(\epsilon'),R\Lambda(\epsilon)]+[\Lambda(\epsilon'),R\Lambda(\epsilon)]\right)\right)\,,
\label{mmcybe}
\end{eqnarray}
which is also proportional to the CYBE. 
Nilpotency on $\omega_{1+}$ is satisfied because
\begin{eqnarray}
\epsilon' Q_{1}(\epsilon Q_{0}(\omega_{1+}))
&=&[S_{2},\epsilon'\lambda_{3}] 
\end{eqnarray}
is proportional to the gauge symmetry generated by the pure spinor constraint \eqref{purespinor}. An analogous result can be obtained for $\omega_{3-}$.
Now, for $\omega^{*}_{1+}$ we find
\begin{eqnarray}
\epsilon' Q_{1}(\epsilon Q_{1}(\omega_{1+}^{*}))
&=&\mathcal{P}_{13}\big(Ad_{g}([R\Lambda(\epsilon'),R\partial_{+}\Lambda(\epsilon)]-R([R\Lambda(\epsilon'),\partial_{+}\Lambda(\epsilon)]+[\Lambda(\epsilon'),R\partial_{+}\Lambda(\epsilon)])\big)_{1}\,,\nonumber
\end{eqnarray}
which is proportional to the CYBE. An analogous result is obtained for $\omega_{3-}^{*}$. Thus, the BRST charge $Q=Q_{0}+\eta Q_{1}$ is nilpotent if $R$ is a solution of the CYBE \cite{Mikhailov2}.

Notice that the $R$-matrix is not constrained  to lie in the kernel of the CYBE. We can clearly see this when considering that the left-hand side of the \eqref{mmcybe} is required to vanish up to a local transformation of $g$ \cite{Berkovits4}. The term $g[\Lambda(\epsilon),\Lambda(\epsilon')]$ is proportional to a Lorentz rotation on $g$, hence, the nilpotency of the BRST charge allows the introduction of  $R$-matrices that solve the mCYBE. This has been extensively discussed in \cite{Bevilaqua}.

Having found $Q_{1}$ the next step is to find $V^{}_{2}$ using \eqref{to}. First of all, we compute the action of $Q_{1}$ on the local currents $\tilde{j}_{\pm}$ 
\begin{eqnarray}
\epsilon Q_{1}\tilde{j}_{-}&=&[\tilde{j}_{-},R\Lambda(\epsilon)]+Ad_{g}^{-1}\circ(d_{PS}-4\mathcal{P}_{31})\circ Ad_{g} \partial_{-}R\Lambda(\epsilon)\,,\\
\epsilon Q_{1}\tilde{j}_{+}&=&[\tilde{j}_{+},R\Lambda(\epsilon)]+Ad_{g}^{-1}\circ(\hat{d}_{PS}-4\mathcal{P}_{13})\circ Ad_{g} \partial_{+}R\Lambda(\epsilon)\,.
\end{eqnarray}
where we introduced $\hat{d}=3P_{1}+2P_{2}+P_{3}$, the transpose operator of $d$. Then, we obtain
\begin{eqnarray}
\epsilon Q_{1}V^{}_{1}&=&
-\frac{1}{4}Str\big([R\tilde{j}_{-},R\tilde{j}_{+},],\Lambda(\epsilon))+\frac{1}{4}Str(\partial_{-}R\Lambda(\epsilon),Ad_{g}^{-1}\circ(d_{PS}-4\mathcal{P}_{31})\circ Ad_{g}R\tilde{j}_{+})+\nonumber\\
&+&\frac{1}{4}Str(\partial_{+}R\Lambda(\epsilon),Ad_{g}^{-1}\circ(\hat{d}_{PS}-4\mathcal{P}_{13})\circ Ad_{g}R\tilde{j}_{-}\big)\,,
\label{contri}
\end{eqnarray}
where we have made use of the CYBE for $R$. It is important to take into account the following identity
\begin{equation}
\epsilon Q_{0}Str(Ad_{g}X,(d_{PS}-4\mathcal{P}_{31})Ad_{g}Y)=Str(\Lambda(\epsilon),[X,Y])\,,
\label{taken}
\end{equation}
valid for any $X$ and $Y$ invariant fields under BRST transformations. Then, it follows that
\begin{eqnarray}
\begin{aligned}
&\epsilon Q_{0}Str(Ad_{g}R\tilde{j}_{+},(d_{PS}-4\mathcal{P}_{31})Ad_{g}R\tilde{j}_{-})\nonumber\\
&=Str([R\tilde{j}_{+},R\tilde{j}_{-}],\Lambda(\epsilon))-Str(\partial_{-}R\Lambda(\epsilon),Ad_{g}^{-1}\circ(d_{PS}-4\mathcal{P}_{31})\circ Ad_{g}R\tilde{j}_{+})\nonumber\\
&+Str(\partial_{+}R\Lambda(\epsilon),Ad_{g}^{-1}\circ(\hat{d}_{PS}-4\mathcal{P}_{13})\circ Ad_{g}R\tilde{j}_{-})\,,
\end{aligned}
\end{eqnarray}
which cancels the contribution from \eqref{contri}. Then, $V_2=\frac{1}{4}Str(Ad_{g}R\tilde{j}_{+},(d_{PS}-4\mathcal{P}_{31})Ad_{g}R\tilde{j}_{-})$ and this solution is consistent if $Q_{2}S_{0}=0$. Since there is no other local fermionic symmetry of the undeformed pure spinor action then it follows that  $Q_{2}=0$.

Having understood this procedure it is not too difficult to find an expression for $V_{3}$. First, we note that 
\begin{eqnarray}
\epsilon Q_{1}Str(Ad_{g}X,(d_{PS}&-&4\mathcal{P}_{31})Ad_{g}Y)=
Str([R\Lambda(\epsilon),X],Ad_{g}^{-1}\circ (\hat{d}_{PS}-4\mathcal{P}_{13})\circ Ad_{g}Y)\nonumber\\
&&+Str([R\Lambda(\epsilon),X],Ad_{g}^{-1}\circ (d_{PS}-4\mathcal{P}_{31})\circ Ad_{g}Y)\,,
\label{qq1}
\end{eqnarray}
which can easily be proven by taking into account the CYBE. Combining \eqref{taken}, \eqref{Qj} and \eqref{qq1} it is possible to show that $Q_{1}V_{2}^{}+Q_{0}V_{3}^{}=0$ if
\begin{eqnarray}
V_{3}^{}=\frac{1}{4}STr\big(\tilde{j}_{+}, R\circ Ad_{g}^{-1}\circ(d_{PS}-4\mathcal{P}_{31})\circ Ad_{g}\circ R \circ Adg^{-1}\circ(d_{PS}-4\mathcal{P}_{31})\circ Ad_{g} R\tilde{j}_{-}  \big
)\, ,\nonumber\\
\end{eqnarray}
so that that $Q_{3}=0$. At this stage it is clear the pattern for the higher order terms.

Then, the complete deformed action, invariant under the BRST charge $Q=Q_{0}+\eta Q_{1}$, takes the form 
\begin{eqnarray}
S_{def}&=&S_{0}-\frac{\eta}{4}Str(\tilde{j}_{+},\mathcal{U}(\eta)R\tilde{j}_{-})\,,\nonumber\\
\mathcal{U}(\eta)&=&\sum_{n=0}^{\infty}(\eta R\circ Ad_{g}^{-1}\circ ({d}_{PS}-4\mathcal{P}_{31})\circ Ad_{g})^{n}\,.
\label{primeradeform}
\end{eqnarray}
We can check this  by splitting the total BRST variation of the action in three parts. First of all we note that 
\begin{equation}
\epsilon QS_{0}=\eta \epsilon Q_{1}S_{0}=\frac{1}{4}[Str(\partial_{-}R\Lambda(\epsilon),R\hat{j}_{+})-Str(\partial_{+}R\Lambda(\epsilon),R\hat{j}_{-})]\,.
\label{342}
\end{equation}
Considering \eqref{taken} we find 
\begin{eqnarray}
\epsilon Q_{0}Str(\hat{j}_{+},\mathcal{U}R\hat{j}_{-})
&=& Str(\partial_{+}\Lambda(\epsilon),\mathcal{U}           R\tilde{j}_{-})+Str(\partial_{-}\Lambda(\epsilon),(\mathcal{U}         R)^{t}\tilde{j}_{+})\nonumber\\
&+&Str(\Lambda(\epsilon),[\mathcal{U}R\tilde{j}_{-},(\mathcal{U}   R)^{t}\tilde{j}_{+}])\,.
\label{346}
\end{eqnarray}
Similarly, taking into account \eqref{qq1}, the action of $ Q_{1}$ over the same expression is given by
\begin{eqnarray}
 \epsilon Q_{1}Str(\tilde{j}_{-},\mathcal{U}R\tilde{j}_{+})
&=&Str([Ad_{g}^{-1}\circ(\hat{d}_{PS}-4\mathcal{P}_{13})\circ Ad_{g} \partial_{+}R\Lambda(\epsilon),\mathcal{U}R\tilde{j}_{-})\nonumber\\
&+& Str(\mathcal{U}R\circ Ad_{g}^{-1}\circ(d_{PS}-4\mathcal{P}_{31})\circ Ad_{g} \partial_{-}R\Lambda(\epsilon))+\nonumber\\
&+& Str([\tilde{j}_{+},R\Lambda(\epsilon)],\mathcal{U}R\tilde{j}_{-}) + Str(\tilde{j}_{+},\mathcal{U}R([\tilde{j}_{-},R\Lambda(\epsilon)])+\nonumber\\
&+&
Str([R\Lambda(\epsilon),(\mathcal{U}R)^{t}\tilde{j}_{+}],Ad_{g}^{-1}\circ (\hat{d}_{PS}-4\mathcal{P}_{13})\circ Ad_{g} \circ \mathcal{U}R\tilde{j}_{-})\nonumber\\
&+&Str([R\Lambda(\epsilon),X],Ad_{g}^{-1}\circ (d_{PS}-4\mathcal{P}_{31})\circ Ad_{g}Y)\,.\nonumber
\end{eqnarray}
When considering the CYBE the above equation is rewritten as
\begin{eqnarray}
 \epsilon Q_{1}Str(\tilde{j}_{+},\mathcal{U}R\tilde{j}_{-})
&=&Str([Ad_{g}^{-1}\circ(\hat{d}_{PS}-4\mathcal{P}_{13})\circ Ad_{g} \partial_{+}R\Lambda(\epsilon),\mathcal{U}R\tilde{j}_{-})+\nonumber\\
&+& Str(\tilde{j}_{+},\mathcal{U}R\circ Ad_{g}^{-1}\circ(d_{PS}-4\mathcal{P}_{31})\circ Ad_{g} \partial_{-}R\Lambda)+\nonumber\\
&+& Str([R\mathcal{U}^{t}\tilde{j}_{+},\mathcal{U}R\tilde{j}_{-}]\Lambda(\epsilon))\,.
\label{323}
\end{eqnarray}
Taking into account \eqref{342}, \eqref{346} and \eqref{323} it is easy to verify that the BRST variation of the complete deformed action vanishes.

\section{\textbf{The polynomial deformed action}}
Recall that the anti-fields were introduced in order to get an off-shell nilpotent BRST charge.  
The deformed action obtained in the last section \eqref{primeradeform} is non polynomial due to the projectors $\mathcal{P}$ which emerged as a consequence of the BRST transformation on the anti-fields \eqref{antg1} and \eqref{antg2}.  
In this section we abandon the anti-fields formulation from the outset and, as a consequence, the advantages of working with an off-shell nilpotent BRST charge. What we get is a local action which is a polynomial expansion of local fields.

First of all we solve \eqref{since} for
\begin{equation}
V^{}_{1}=\frac{1}{4}\int Str(Rj_{+},j_{-})\,.
\label{hatjj1}
\end{equation}
Taking into account \eqref{Qj} we have that
\begin{eqnarray}
\epsilon Q_{0}V_{1}^{}&=&-\frac{1}{4}(Str(j_{+},\partial_{-}R\Lambda(\epsilon))-Str(j_{-},\partial_{+}R\Lambda(\epsilon)))-\label{kkk}\\
&-&Str(D_{+}\epsilon\lambda_{1}
-[N_{0-},\epsilon\lambda_{1}],Ad_{g}\circ Rj_{-})
+Str(D_{-}\epsilon\lambda_{3}-[N_{0+},\epsilon\lambda_{3}],Ad_{g}\circ Rj_{+})\,.\nonumber
\end{eqnarray}
We look for a BRST operator $Q_{1}$ consistent with the PS formulation \cite{BH} so this requires $Q_{1}(\lambda)=0$. Our ansatz then takes the form
\begin{equation}
Q_{1}=gR\Lambda(\epsilon)\frac{\delta}{\delta g}+\alpha_{1+}\frac{\delta}{\delta \omega_{1+}}+\beta_{3-} \frac{\delta}{\delta \omega_{3-}}\,,
\label{q1}
\end{equation}
where $\alpha$ and $\beta$ are to be fixed by imposing BRST invariance at first order in $\eta$. Its action over $S_{0}$ is given by
\begin{eqnarray}
\epsilon Q_{1}S_{0}=-\big( Str(D_{+}\epsilon\lambda_{1}
-[N_{0-},\epsilon\lambda_{1}],\beta_{3-})-Str(D_{-}\epsilon\lambda_{3}-[N_{0+},\epsilon\lambda_{3}],\alpha_{1+})\big)+\nonumber\\
+\frac{1}{4}\big(Str(j_{+},\partial_{-}R\Lambda(\epsilon))-Str(j_{-},\partial_{+}R\Lambda(\epsilon))\big)\,.
\end{eqnarray}
Then, to cancel \eqref{kkk} the action of $Q_{1}$ on the ghost sector must be
\begin{equation}
\epsilon Q_{1}(\omega_{1+})=-P_{1}(\epsilon Ad_{g}\circ Rj_{+})\,,\qquad \epsilon Q_{1}(\omega_{3-})=-P_{3}(\epsilon Ad_{g}\circ Rj_{-})\,.
\end{equation}

Having obtained $Q_{1}$ we look for $V^{}_{2}$ by solving \eqref{to}. Let us now write the $Q_{1}$  transformations of $j_{\pm}$ 
\begin{eqnarray}
\epsilon Q_{1}j_{+}=[j_{+},R\Lambda(\epsilon)]+Ad_{g}^{-1}\circ\hat{d}_{PS}\circ Ad_{g} \partial_{+}R\Lambda(\epsilon)-4Ad_{g}^{-1}[\epsilon\lambda_{3-},P_{1}\circ Ad_{g} Rj_{+}]\,,\\
\epsilon Q_{1}j_{-}=[j_{-},R\Lambda(\epsilon)]-Ad_{g}^{-1}\circ\ d_{PS}\circ Ad_{g}\partial_{-}R\Lambda(\epsilon)+4Ad_{g}^{-1}[\epsilon\lambda_{1+},P_{3}\circ Ad_{g} Rj_{-}]\,.
\end{eqnarray}
This means that
\begin{eqnarray}
\begin{aligned}
&\epsilon Q_{1}V^{}_{1}=\label{olvidado}\\
&-\frac{1}{4}Str([Rj_{-},j_{+}],R\Lambda(\epsilon))-\frac{1}{4}Str(\partial_{+}R\Lambda(\epsilon),Ad_{g}^{-1}\circ\ d_{PS}\circ Ad_{g} R j_{-})\\
&+Str([\epsilon\lambda_{3-},(Ad_{g}Rj_{+})_{1}],Ad_{g}\circ Rj_{-})
+\frac{1}{4}Str([Rj_{+},j_{-}],R\Lambda(\epsilon))\\
&-\frac{1}{4}Str(\partial_{-}R\Lambda(\epsilon),Ad_{g}^{-1}\circ\hat{d}_{PS}\circ Ad_{g}  Rj_{+})+Str([\epsilon\lambda_{1+},(Ad_{g} Rj_{-})_{3}],Ad_{g}\circ Rj_{+})\,.
\end{aligned}
\end{eqnarray}
The natural ansatz for $V^{}_{2}$ is then $\frac{1}{4}Str(j_{+},R\circ Ad_{g}^{-1}\circ d_{PS}\circ Ad_{g}\circ Rj_{+})$. In order to show this we need to find a $Q_{2}$ such that \eqref{to} holds. 
First, we observe that 
\begin{eqnarray}
\epsilon Q_{L0}Str(Ad_{g}X, d_{PS}\circ Ad_{g}Y)
&=&Str(\epsilon\lambda_{3-},[gXg^{-1}, gYg^{-1}]-4[(gXg^{-1})_{1}, (gYg^{-1})_{0}])\,,\nonumber\\
\epsilon Q_{R0}Str(Ad_{g}X, d_{PS}\circ Ad_{g}Y)
&=&-Str(\epsilon\lambda_{1+},[gXg^{-1}, gYg^{-1}]+4[(gXg^{-1})_{0}, (gYg^{-1})_{3}])\,.\nonumber\\
\end{eqnarray}
These results altogether give
\begin{eqnarray}
\begin{aligned}
&\epsilon Q_{0}Str(Ad_{g}X, d_{PS}\circ Ad_{g}Y)=-Str(\Lambda(\epsilon),[X,Y])\\
&-4Str(\epsilon\lambda_{1+},[(gXg^{-1})_{0}, (gYg^{-1})_{3}])-4Str(\epsilon\lambda_{3-},[(gXg^{-1})_{1}, (gYg^{-1})_{0}])\,,
\end{aligned}
\end{eqnarray}
and we find
\begin{eqnarray}
\begin{aligned}
&\epsilon Q_{0}Str(Ad_{g}\circ Rj_{+}, d_{PS}\circ Ad_{g}\circ Rj_{-})=-Str(\Lambda(\epsilon),[Rj_{+},Rj_{-}])\\
&-4(Str(\epsilon\lambda_{1+},[(gRj_{+}g^{-1})_{0}, (gRj_{-}g^{-1})_{3}])+Str(\epsilon\lambda_{3-},[(gRj_{+}g^{-1})_{1}, (gRj_{-}g^{-1})_{0}]))\\
&-Str(\partial_{+}R\Lambda(\epsilon),Ad_{g}^{-1}\circ d_{PS}\circ Ad_{g}Rj_{-})-Str(\partial_{-}R\Lambda(\epsilon),Ad_{g}^{-1}\circ \hat{d}_{PS}\circ Ad_{g}Rj_{+})+\\
&+4Str(R_{g}\circ \hat{d}_{PS}\circ Ad_{g}\circ Rj_{+},D_{-}\lambda_{3}-[N_{0+},\epsilon\lambda_{3}])\\
&-4Str(R_{g}\circ d_{PS}\circ Ad_{g}\circ Rj_{-},D_{-}\epsilon\lambda_{1}-[N_{0+},\lambda_{1}])\,,
\label{h}
\end{aligned}
\end{eqnarray}
where $ R_{g}=Ad_{g}\circ R \circ Ad_{g}^{-1}$. 
The last two terms of \eqref{h} are proportional to the equations of motion of $\omega_{1}$ and $\omega_{3}$ and can be removed by $Q_{2}S_{0}$ if $Q_{2}$ is defined as
\begin{eqnarray}
Q_{2}(\omega_{1+})&=&P_{1}(Ad_{g}\circ R\circ Ad_{g}^{-1}\circ d_{PS}\circ Ad_{g}\circ Rj_{+})\,,\\  Q_{2}(\omega_{3-})&=&-P_{3}(Ad_{g}\circ R\circ Ad_{g}^{-1}\circ \hat{d}_{PS}\circ Ad_{g}\circ Rj_{-})\,.
\end{eqnarray}
The remaining terms in \eqref{h} cancel with the contribution coming from \eqref{olvidado}.
This result allows us to infer the pattern of deformation for the higher order terms.

Taking into account the above results the complete deformed action is given by the local expansion
\begin{equation}
S_{def}=S-\frac{\eta}{4}Str(j_{+},\tilde{\mathcal{U}}(\eta)Rj_{-})\,,\qquad \tilde{\mathcal{U}}(\eta)=\sum_{n=0}^{\infty}(\eta R\circ Ad_{g}^{-1}\circ d_{PS}\circ Ad_{g})^{n}\,.
\end{equation}
We can rearrange this action into a more familiar form by defining the operators
\begin{equation}
\mathcal{O}_{PS-}=1-\eta R_{g}d_{PS}\,,\quad \mathcal{O}_{PS+}=1+\eta R_{g}\hat{d}_{PS}\,.
\label{aclaracion}
\end{equation}
Also, it is useful to consider the deformed pure spinor currents
 \begin{equation}
 \bar{J}_{\pm}=-\mathcal{O}^{-1}_{PS\pm}\partial_{+}g\,g^{-1}\,,\qquad J_{\pm}=-\mathcal{O}^{-1}_{PS\pm}\partial_{-}g\,g^{-1}. \label{416}
 \end{equation}
Hence, the deformed action can be rewritten as \footnote{We introduce hatted and unhatted currents in order to avoid confusion between the $\pm$ indices in \eqref{aclaracion} and the light-cone coordinates indices.}
 \begin{eqnarray}
 S_{def}&=&\int\big[ \frac{1}{4}Str(\partial_{+}g\,g^{-1},d_{PS}J_{-})+Str(N_{0+}J_{0-}+N_{0-}\bar{J}_{0+})+\nonumber\\
 &-&Str(N_{0-}(1-4\eta\mathcal{O}^{-1}_{PS-}R_{g})N_{0+})+Str(\omega_{1+}\partial_{-}\lambda_{3}+\omega_{3-}\partial_{+}\lambda_{1})\big]\,.
 \label{deformado}
 \end{eqnarray}

This action must be invariant under the BRST transformations
\begin{eqnarray}
\epsilon Q(g)&=&\{(1-\eta R_{g})\epsilon\lambda_{1}+(1+\eta R_{g})\epsilon\lambda_{3}\}g\,,\\
Q(w_{3-})&=&-J_{3-}-4\eta P_{3}\circ\mathcal{O}^{-1}_{PS-}R_{g}N_{0-}\,,\label{brst-}\\
Q(w_{1+})&=&-\bar{J}_{1+}+4\eta P_{1}\circ\mathcal{O}^{-1}_{PS+}R_{g}N_{0+}\,.
\label{brst+}
\end{eqnarray}
This can be shown by splitting  the action into four sectors. Taking into account that the deformed currents vary under $\delta g=g\xi_{i}$ as
\begin{eqnarray}
\delta J_{\pm}&=&-\mathcal{O}^{-1}_{PS\pm}\left(d\xi+[ J_{\pm},\xi] \mp \eta R_{g}[\xi,d_{PS}J_{\pm}]\right)\,,
\label{varJ}
\end{eqnarray} 
the first term of \eqref{deformado}, the matter sector, contributes with
\begin{equation}
\int Str(\delta g g^{-1},\mathcal{E}_{0})\,,
\label{materia}
\end{equation}
\begin{equation}
\mathcal{E}_{0}=\partial_{+}(d_{PS}J_{-})+\partial_{-}(\hat{d}_{PS}\bar{J}_{+})+[\bar{J}_{+},dJ_{-}]+[J_{-},\hat{d}\bar{J}_{+}]\,.
\label{mot}
\end{equation}
In particular, for $\delta g g^{-1}=\epsilon Q(g)g^{-1}=(1-\eta R_{g})\epsilon\lambda_{1}+(1+\eta R_{g})\epsilon\lambda_{3}$, and taking into account the following identities
\begin{eqnarray}
P_{1}\circ (1-\eta R_{g})(\mathcal{E}_{0})&=&-4\tilde{D}_{+}J_{3-}\,,\qquad \tilde{D}_{+}=\partial_{+}+[\bar{J}_{0+},\,]\,,\\
P_{3}\circ (1+\eta R_{g})(\mathcal{E}_{0})&=&-4\tilde{D}_{-}\bar{J}_{1+}\,,\qquad \tilde{D}_{-}=\partial_{-}+[J_{0-},\,]\,,
\label{identidad}
\end{eqnarray}
we combine \eqref{identidad} and \eqref{materia} to obtain the BRST transformation of the matter sector 
\begin{equation}
-\int \left( Str(\epsilon\lambda_{1},\tilde{D}_{0+}J_{3-})+Str(\epsilon\lambda_{3},\tilde{D}_{0-}\bar{J}_{1+}\right)\,.
\label{777}
\end{equation}

For the matter-ghost sector, the second term in \eqref{deformado}, we note that the BRST transformations of $J_{-}$ and $\bar{J}_{+}$ are
\begin{eqnarray}
\epsilon Q(J_{-})&=& \mathcal{O}^{-1}_{PS-}\left[[\epsilon Qg g^{-1},J_{-}]-\eta R_{g}[\epsilon Qg g^{-1},d_{PS}J_{-}]-\partial_{-}(\epsilon Qg g^{-1})\right]\,,\\
\epsilon Q (\bar{J}_{+})&=& \mathcal{O}^{-1}_{PS+}\left[[\epsilon Qg g^{-1},\bar{J}_{+}]-\eta R_{g}[\epsilon Qg g^{-1},\hat{d}_{PS}\bar{J}_{+}]-\partial_{+}(\epsilon Qg g^{-1})\right]\,.
\end{eqnarray}
After a lengthy calculation we obtain that
\begin{eqnarray}
Str(\epsilon Q (J_{0-}),N_{0+})&=&Str(\epsilon \lambda_{1}, [J_{3-},N_{0+}])-4Str(\epsilon \lambda_{3}, \partial_{-}( \mathcal{O}^{-1}_{PS+}R_{g}N_{0+}])+\\
&+&4\eta Str(\epsilon \lambda_{1}, [J_{3-},\mathcal{O}^{-1}_{PS+}R_{g}N_{0+}])-4\eta Str(\epsilon \lambda_{3}, [J_{0-},\mathcal{O}^{-1}_{PS+}R_{g}N_{0+}])\,,\nonumber\\
Str(\epsilon Q (J_{0+}),N_{0-})&=&Str(\epsilon \lambda_{3}, [\bar{J}_{1+},N_{0-}])-4Str(\epsilon \lambda_{1}, \partial_{+}( \mathcal{O}^{-1}_{PS-}R_{g}N_{0-}])+\label{666}
\\
&+&4\eta Str(\epsilon \lambda_{3}, [\bar{J}_{1+},\mathcal{O}^{-1}_{PS-}R_{g}N_{0-}])-4\eta Str(\epsilon \lambda_{1}, [\bar{J}_{0+},\mathcal{O}^{-1}_{PS-}R_{g}N_{0-}])\,.\nonumber
\end{eqnarray}
On the other hand, considering \eqref{brst-} and \eqref{brst+}, we have 
\begin{eqnarray}
\epsilon Q(N_{0-})&=&\{J_{3-}+4\eta P_{3}\circ\mathcal{O}^{-1}_{PS-}R_{g}N_{0-},\epsilon\lambda_{3}\}\,,\\
\epsilon Q(N_{0+})&=&\{\bar{J}_{1+}+4\eta P_{1}\circ\mathcal{O}^{-1}_{PS+}R_{g}N_{0+},\epsilon\lambda_{1}\}\,,
\label{transN}
\end{eqnarray}
so that 
\begin{eqnarray}
Str(J_{0-},\epsilon QN_{0+})&=&Str(\epsilon \lambda_{3}, [\bar{J}_{1+},J_{0-}])-Str(\epsilon \lambda_{3},\eta[P_{1}\circ\mathcal{O}^{-1}_{PS+}R_{g}N_{0+},J_{0-}])\,,\label{5555}
\\
Str(\bar{J}_{0+},\epsilon QN_{0-})&=&Str(\epsilon \lambda_{1}, [J_{3-},\bar{J}_{0+}])-Str(\epsilon \lambda_{1},\eta[P_{3}\circ\mathcal{O}^{-1}_{PS-}R_{g}N_{0-},\bar{J}_{0+}])\,.
\label{555}
\end{eqnarray}

In the ghost sector we have to consider \eqref{brst-} and \eqref{brst+} to set
\begin{eqnarray}
\epsilon Q\,S_{gh}&=&Str(\epsilon \lambda_{1}, \partial_{+}J_{3-})+Str(\epsilon \lambda_{1},4\eta P_{3}\circ\partial_{+}(\mathcal{O}^{-1}_{PS-}R_{g}N_{0-}))\,,\nonumber\\
&+&Str(\epsilon \lambda_{3}, \partial_{-}\bar{J}_{1+})+Str(\epsilon \lambda_{3},4\eta P_{1}\circ\partial_{-}(\mathcal{O}^{-1}_{PS+}R_{g}N_{0+}))\,.
\label{444}
\end{eqnarray}
For the third term of \eqref{deformado} we have 
\begin{eqnarray}
&&\epsilon Q Str(N_{0+},(1-4\eta\mathcal{O}^{-1}_{PS-}R_{g})N_{0-})=\label{finally}\\
&& Str(\epsilon QN_{0+},(1-4\eta\mathcal{O}^{-1}_{PS-}R_{g})N_{0-})+Str(N_{0+},(1-4\eta\mathcal{O}^{-1}_{PS-}R_{g})\epsilon QN_{0+})+\nonumber\\
&&-4\eta Str(N_{0-},\epsilon Q(\mathcal{O}^{-1}_{PS-}R_{g})\circ N_{0-})\,.\nonumber
\end{eqnarray}
Considering \eqref{transN}, the first term of the above equation can be expressed as
\begin{eqnarray}
&-&Str(\epsilon \lambda_{3},[N_{0-},\bar{J}_{1+}])+4\eta Str(\epsilon \lambda_{3},[N_{0-},P_{1}\circ\mathcal{O}^{-1}_{PS+}R_{g}N_{0+}])+\label{111}\\
&-&4\eta Str(\epsilon \lambda_{3},[P_{0}\circ\mathcal{O}^{-1}_{PS-}R_{g}N_{0-},\bar{J}_{1+}])+\nonumber\\
&+&4\eta^{2}Str(\epsilon \lambda_{3},[P_{0}\circ\mathcal{O}^{-1}_{PS-}R_{g}N_{0-},P_{1}\circ\mathcal{O}^{-1}_{PS+}R_{g}N_{0+}])\,,\nonumber
\end{eqnarray}
while the second one is given by
\begin{eqnarray}
&-&Str(\epsilon \lambda_{1},[N_{0+},J_{3-}])+4\eta Str(\epsilon \lambda_{1},[N_{0+},P_{3}\circ\mathcal{O}^{-1}_{PS-}R_{g}N_{0-}])+\label{222}\\
&-&4\eta Str(\epsilon \lambda_{1},[P_{0}\circ\mathcal{O}^{-1}_{PS+}R_{g}N_{0+},J_{3-}])+\nonumber\\
&+&4\eta^{2}Str(\epsilon \lambda_{1},[P_{0}\circ\mathcal{O}^{-1}_{PS+}R_{g}N_{0+},P_{3}\circ\mathcal{O}^{-1}_{PS-}R_{g}N_{0-}])\,.\nonumber
\end{eqnarray}
After a lengthly computation the last term in \eqref{finally} can be expressed as
\begin{eqnarray}
&-&4\eta Str(\epsilon \lambda_{3},[N_{0-},P_{1}\circ\mathcal{O}^{-1}_{PS+}R_{g}N_{0+}])-4\eta Str(\epsilon \lambda_{1},[N_{0+},P_{3}\circ\mathcal{O}^{-1}_{PS-}R_{g}N_{0-}])+
\label{333}\\
&-&4\eta^{2}Str(\epsilon \lambda_{1},[P_{0}\circ\mathcal{O}^{-1}_{PS+}R_{g}N_{0+},P_{3}\circ\mathcal{O}^{-1}_{PS-}R_{g}N_{0-}])-\nonumber\\
&-&4\eta^{2}Str(\epsilon \lambda_{3},[P_{0}\circ\mathcal{O}^{-1}_{PS-}R_{g}N_{0-},P_{1}\circ\mathcal{O}^{-1}_{PS+}R_{g}N_{0+}])\nonumber\,.
\end{eqnarray}
The contributions \eqref{777}, \eqref{666}, \eqref{5555}, \eqref{555}, \eqref{444}, \eqref{111}, \eqref{222}, and \eqref{333} vanish showing that the action is BRST invariant.

\section{Integrability}
To derive the equations of motion from \eqref{deformado} it is convenient to split the action in three sectors. For the matter sector we implement the variation in the form $\delta g=g\xi_{i}$ to get a contribution as in \eqref{mot}. 
For the matter-ghost sector we have to consider 
\begin{eqnarray}
&\int& ( Str( \delta J_{0-},N_{0+})+Str( \delta \bar{J}_{0+},N_{0-}) )=
\int Str(\delta g g^{-1},\mathcal{E}_{1})\,,
\label{defF}
\end{eqnarray}
where
\begin{eqnarray}
\mathcal{E}_{1}&=&\partial_{-}(\hat{d}_{PS}\mathcal{O}^{-1}_{PS+}R_{g}N_{0+})+[J_{-},N_{0+}]-4\eta[J_{-},\hat{d}\mathcal{O}^{-1}_{PS+}R_{g}N_{0+}]-[\mathcal{O}^{-1}_{PS+}R_{g}N_{0+},dJ_{-}]+\nonumber\\
&+&\partial_{+}(d_{PS}\mathcal{O}^{-1}_{PS-}R_{g}N_{0-})+[\bar{J}_{+},N_{0-}]+4\eta[\bar{J}_{+},d\mathcal{O}^{-1}_{PS-}R_{g}N_{0-}]+[\mathcal{O}^{-1}_{PS-}R_{g}N_{0-},\hat{d}\bar{J}_{+}]\,.\nonumber
\end{eqnarray}
In the $N_{0-}N_{0+}$ sector we have 
\begin{equation}
 Str(N_{0+},\delta(\mathcal{O}^{-1}_{PS-}R_{g})N_{0-})=Str\big(\delta g g^{-1},\mathcal{E}_{2}\big)\,,
\label{varN}
\end{equation}
where
\begin{eqnarray}
\mathcal{E}_{2}&=&[N_{0+}, \mathcal{O}^{-1}_{PS-}R_{g}N_{0-}]+[\mathcal{O}^{-1}_{PS+}R_{g}N_{0+},N_{0-}]+\nonumber\\
&+&[\hat{d}\mathcal{O}^{-1}_{PS+}R_{g}N_{0+}, \mathcal{O}^{-1}_{PS-}R_{g}N_{0-}]+
[\mathcal{O}^{-1}_{PS+}R_{g}N_{0+},d\mathcal{O}^{-1}_{PS-}R_{g}N_{0-}]\,.
\label{varNN}
\end{eqnarray}
Collecting all contributions the equations of motion take the form
\begin{eqnarray}
{\mathcal{E}}&\equiv&
\partial_{+}d_{PS}(J_{-}+4\eta \mathcal{O}^{-1}_{PS-}R_{g}N_{0-})+\partial_{-}\hat{d}_{PS}(\bar{J}_{+}-4\eta \mathcal{O}^{-1}_{PS+}R_{g}N_{0+})+\nonumber\\
&+&[(\bar{J}_{+}-4\eta \mathcal{O}^{-1}_{PS+}R_{g}N_{0+}),d(J_{-}+4\eta \mathcal{O}^{-1}_{PS-}R_{g}N_{0-})]
+\nonumber\\
&+&[(J_{-}+4\eta \mathcal{O}^{-1}_{PS-}R_{g}N_{0-}),\hat{d}(\bar{J}_{+}-4\eta \mathcal{O}^{-1}_{PS+}R_{g}N_{0+})]+\,.\nonumber\\
&+&[(J_{-}+4\eta \mathcal{O}^{-1}_{PS-}R_{g}N_{0-}),N_{0+}]
+[(\bar{J}_{+}-4\eta \mathcal{O}^{-1}_{PS+}R_{g}N_{0+}),N_{0-}]=0\,.
\label{eq}
\end{eqnarray}

For the ghost currents the equations of motion can be expressed as
\begin{eqnarray}
\mathcal{G}_{1}\equiv
\partial_{-}N_{0+}+[(J_{-}+4\eta \mathcal{O}^{-1}_{PS-}R_{g}N_{0-}),N_{0+}]-[N_{0-},N_{0+}]&=&0\,,\label{g1}\\
\mathcal{G}_{2}\equiv
\partial_{+}N_{0-}+[(\bar{J}_{+}-4\eta \mathcal{O}^{-1}_{PS+}R_{g}N_{0+}),N_{0-}]-[N_{0+},N_{0-}]&=&0\,\label.
\label{eq2}
\end{eqnarray}
We notice that the equations of motion \eqref{eq}-\eqref{eq2} present the same structure as the  undeformed ones as expected \cite{vicedo,delduc,vanTongeren:2018vpb}. 
We can write them in a more suggestive manner by defining the currents $\mathcal{J}_{+}$ and $\mathcal{J}_{-}$ as
\begin{equation}
\mathcal{J}_{\pm}=\bar{J}_{\pm} \mp 4\eta \mathcal{O}^{-1}_{PS\pm}R_{g}N_{0\pm}\,,
\label{defjotas}
\end{equation}
so that the equation of motion \eqref{eq} takes the form 
\begin{eqnarray}
\mathcal{E}=
\partial_{+}(d_{PS}\mathcal{J}_{-})+\partial_{-}(\bar{d}_{PS}\mathcal{J}_{+})+[\mathcal{J}_{+},d\mathcal{J}_{-}]+[\mathcal{J}_{+},\hat{d}\mathcal{J}_{-}]+[\mathcal{J}_{-},N_{0+}]+[\mathcal{J}_{+},N_{0-}]=0\,.\nonumber
\label{H}
\end{eqnarray}
and for the ghost currents (\ref{g1}) and (\ref{eq2}),
\begin{eqnarray}  
\mathcal{G}_{1}=
\partial_{-}N_{0+}+[\mathcal{J}_{0-},N_{0+}]-[N_{0-},N_{0+}]&=&0\,,\\
\mathcal{G}_{2}=
\partial_{+}N_{0-}+[\mathcal{J}_{0+},N_{0-}]-[N_{0+},N_{0-}]&=&0\,.
\label{G}
\end{eqnarray}

At this stage it should be clear that the ansatz for the Lax pair can be found by exchanging $\mathcal{J}_{-}$ and $\mathcal{J}_{+}$ for $J_{}$ and $\bar{J}_{+}$, respectively, in the undeformed Lax pair \eqref{laxundef}. This is expected since  the pair of currents ($\mathcal{J}_{-}$, $\mathcal{J}_{+}$) satisfy the zero curvature condition when the classical equations of motion are imposed.
This can be shown by inverting \eqref{defjotas} as
\begin{eqnarray}
A_{\pm}=\mathcal{O}_{PS\pm}\mathcal{J}_{\pm}\mp 4\eta R_{g}N_{0\pm}\,.
\end{eqnarray}
Each term in the Maurer-Cartan equation
$
\partial_{-}A_{+}-\partial_{+}A_{-}+[A_{-},A_{+}]=0$  can be expressed as
\begin{align}
\partial_{-}A_{+}&=\partial_{-}\mathcal{J}_{+}+[A_{+}-\mathcal{J}_{+},A_{-}]  +\eta  R_{g}\big(\partial_{-}(\hat{d}_{PS}\mathcal{J}_{+})-\partial_{-}N_{0+}+[A_{-},\hat{d}_{PS}\mathcal{J}_{+}-N_{0+}]\big)\,,\nonumber\\
\partial_{+}A_{-}&=\partial_{+}\mathcal{J}_{-}+[A_{-}-\mathcal{J}_{-},A_{+}] -\eta  R_{g}\big(\partial_{+}(d_{PS}\mathcal{J}_{-})-\partial_{+}N_{0-}+[A_{+},d_{PS}\mathcal{J}_{-}-N_{0-}]\big)\,,\nonumber\\
[A_{-},A_{+}]&=[\mathcal{J}_{-},\mathcal{J}_{+}]+[\mathcal{J}_{-},A_{+}-\mathcal{J}_{+}]+[A_{-}-\mathcal{J}_{-},\mathcal{J}_{+}]-
\eta^{2}\big([R_{g}d_{PS}\mathcal{J}_{-},R_{g}\hat{d}_{PS}\mathcal{J}_{+}]-\nonumber\\
&-[R_{g}d_{PS}\mathcal{J}_{-},R_{g}N_{0+}]
-[R_{g}\hat{d}_{PS}\mathcal{J}_{+},R_{g}N_{0-}]-[R_{g}N_{0+},R_{g}N_{0-}] \big)\,,
\label{laxtres}
\end{align}
so that, after a lengthy calculation, the Maurer-Cartan equation takes the form 
\begin{eqnarray}
\partial_{-}\mathcal{J}_{+}-\partial_{+}\mathcal{J}_{-}+[\mathcal{J}_{-},\mathcal{J}_{+}]+\eta R_{g}(\mathcal{E})
-\eta R_{g}(\mathcal{G}_{1}+\mathcal{G}_{1})=0\,,
\end{eqnarray}
which shows that the pair $(\mathcal{J}_{-}$, $\mathcal{J}_{+})$ satisfies the zero curvature condition when the equations of motion hold. 

Defining the covariant derivatives as
\begin{equation}
\mathcal{D}_{\pm}=\partial_{\pm}+[\mathcal{J}_{0\pm},\,\,]\,, 
\end{equation}
the equations of motion can be written as
\begin{eqnarray}
\mathcal{D}_{-}\mathcal{J}_{1+}&+&[\mathcal{J}_{1-},N_{0+}]-[N_{0-},\mathcal{J}_{1+}]=0\,,\\
\label{desdedefor}
\mathcal{D}_{-}\mathcal{J}_{2+}&+&[\mathcal{J}_{1-},\mathcal{J}_{1+}]+[\mathcal{J}_{2-},N_{0+}]-[N_{0-},\mathcal{J}_{2+}]=0\,,\\
\mathcal{D}_{-}\mathcal{J}_{3+}&+&[\mathcal{J}_{1-},\mathcal{J}_{2+}]+[\mathcal{J}_{2-},\mathcal{J}_{1+}]-[\mathcal{J}_{3-},N_{0-}]-[N_{0-},\mathcal{J}_{3+}]=0\,,\\
\mathcal{D}_{+}\mathcal{J}_{1-}&+&[\mathcal{J}_{2+},\mathcal{J}_{3-}]+[\mathcal{J}_{3+},\mathcal{J}_{2-}]+
[\mathcal{J}_{1-},N_{0+}]-[N_{0-},\mathcal{J}_{1+}]=0\,,\\
\mathcal{D}_{+}\mathcal{J}_{2-}&+&[\mathcal{J}_{3+},\mathcal{J}_{3-}]+[\mathcal{J}_{2-},N_{0+}]-[N_{0-},\mathcal{J}_{2+}]=0\,,\\
\mathcal{D}_{+}\mathcal{J}_{3-}&+&[\mathcal{J}_{3-},N_{0+}]-[N_{0-},\mathcal{J}_{3+}]=0\,.
\label{hastadefor}
\end{eqnarray}
Similarly, the equations of motion for the ghost sector in terms of the Lorentz currents are
\begin{equation}
\mathcal{D}_{\pm}N_{0\mp}-[N_{0\pm},N_{0\mp}]=0\,.
\label{Necdef}
\end{equation}

We have then shown that the equations of motion for the deformed action admits a zero-curvature representation given by the Lax pair:
\begin{eqnarray}
\mathcal{L}_{+}(z)&=&\mathcal{J}_{0+}+z^{-3}\mathcal{J}_{1+}+z^{-2}\mathcal{J}_{2+}+z^{-1}\mathcal{J}_{3+}+(z^{-4}-1)N_{0+}\,,\\
\mathcal{L}_{-}(z)&=&\mathcal{J}_{0-}+z \mathcal{J}_{1-}+z^{2}\mathcal{J}_{2-}+z^{3}\mathcal{J}_{3-}+(z^{4}-1)N_{0-}\,,
\end{eqnarray}
where $z$ is the spectral parameter. An interesting property of the matter sector is that the equations of motion accept a Lax representation as can be seen by switching off the ghost contributions.

Considering \eqref{brst-} and \eqref{brst+} the BRST  charges are given by  
\begin{equation}
Q^{}_{-}=\oint Str(\lambda_{1},\mathcal{J}_{3-})\,, \qquad Q^{}_{+}=\oint Str(\lambda_{3},\mathcal{J}_{1+})\,.
\end{equation}
The (anti) holomorphicity of the BRST currents can be easily proven by using the above equations of motion.

\section{Relation to the homogeneous YB deformations of the GS superstring}
In this section we will look for the background fields of the deformed pure spinor action. This is achieved by comparing the deformed model \eqref{deformado} with the standard Berkovits-Howe action \cite{BH}
\begin{eqnarray}
S_{BH}&=&\frac{1}{2\pi\alpha'}\int dz^{2}\big(\frac{1}{2}E^{a}\bar{E}^{b}\eta_{ab}+\frac{1}{2}E^{A}\bar{E}^{B}B_{AB}+d_{\alpha}\bar{E}^{\alpha}+d_{\hat{\alpha}}E^{\hat{\alpha}}+d_{\alpha}d_{\hat{\alpha}}P^{\alpha\hat{\alpha}}+\nonumber\\
&+&\Omega_{\alpha}^{\beta}\lambda^{\alpha}\omega_{\beta}+ \hat{\Omega}_{\hat{\alpha}}^{\hat{\beta}}\hat{\lambda}^{\hat{\alpha}}\hat{\omega}_{\hat{\beta}}+\lambda^{\alpha}\omega_{\beta}\hat{d}_{\hat{\gamma}}C_{\alpha}^{\beta\hat{\gamma}}+\hat{\lambda}^{\hat{\alpha}}\hat{\omega}_{\hat{\beta}}d_{\gamma}\tilde{C}_{\hat{\alpha}}^{\hat{\beta}\gamma}+
\lambda^{\alpha}\omega_{\beta}\hat{\lambda}^{\hat{\alpha}}\hat{\omega}_{\hat{\beta}}S^{\beta\hat{\beta}}_{\alpha\hat{\alpha}}+S_{gh}\big)
\,.
\label{BH}
\end{eqnarray}
This is the most general action that possesses BRST symmetry, classical world-sheet conformal invariance and zero ghost number.
Here, $E^{A}$ and ($\Omega_{\alpha}{}^{\beta}, \hat{\Omega}_{\hat{\alpha}}{}^{\hat{\beta}})$ 
 are the super-vielbiens, and the left and right-moving spin connection, and $A=(a,\alpha,\hat{\alpha})$ is a tangent space index. The action also includes the ghosts $(\lambda^{\alpha}, \omega_{\beta},\hat{\lambda}^{\hat{\alpha}}, \hat{\omega}_{\hat{\beta}})$ and the world-sheet auxiliary fields $(d_{\alpha}, d_{\hat{\alpha}})$. These world-sheet fields are coupled through target space fields. The superfield $B_{AB}$ is a superspace two-form. The leading component of $P^{\alpha\hat{\alpha}}$ is the Ramond-Ramond bispinor. The $(C_{\alpha}^{\beta\hat{\gamma}},\tilde{C}_{\hat{\alpha}}^{\hat{\beta}\gamma})$ are related to the gravitini and dilatini, and  $ S^{\beta\hat{\beta}}_{\alpha\hat{\alpha}}$ is related to the Riemann curvature.
As stated above, the pair $(d_{\alpha}, d_{\hat{\alpha}})$ are auxiliary fields and can be integrated out when $P^{\alpha\hat{\alpha}}$ is invertible. Defining its inverse as $P_{\alpha\hat{\alpha}}P^{\beta\hat{\alpha}}=\delta^{\beta}_{\alpha}$, the equations of motion for $d_{\alpha}$ and $d_{\hat{\alpha}}$  give us
\begin{eqnarray}
d_{\alpha}&=&P_{\alpha\hat{\alpha}}(E^{\hat{\alpha}}+\lambda^{\rho}\omega_{\beta}C_{\rho}^{\beta\hat{\alpha}})\,,\\
\hat{d}_{\hat{\alpha}}&=&-P_{\alpha\hat{\alpha}}(\bar{E}^{\alpha}+\hat{\lambda}^{\hat{\rho}}\hat{\omega}_{\hat{\beta}}\tilde{C}_{\hat{\rho}}^{\hat{\beta}\alpha})\,.
\label{d}
\end{eqnarray}
Substituting these values in \eqref{BH} the action takes the form 
\begin{eqnarray}
S_{BH}&=&\frac{1}{2\pi\alpha'}\int dz^{2}\big[\frac{1}{2}E^{a}\bar{E}^{b}\eta_{ab}+\frac{1}{2}E^{A}\bar{E}^{B}B_{AB}-\bar{E}^{\alpha}P_{\alpha\hat{\alpha}}E^{\hat{\alpha}}+\label{integrada}\\
&+&\lambda^{\alpha}\omega_{\beta}(\Omega_{\alpha}^{\beta}-P_{\alpha\hat{\alpha}}C_{\alpha}^{\beta\hat{\alpha}}\bar{E}^{\alpha})+\hat{\lambda}^{\hat{\alpha}\hat{\omega}}_{\hat{\beta}}(\hat{\Omega}_{\hat{\alpha}}^{\hat{\beta}}-P_{\alpha\hat{\alpha}}\tilde{C}_{\hat{\alpha}}^{\hat{\beta}\alpha}E^{\hat{\alpha}})
+\lambda^{\alpha}\omega_{\beta}\hat{\lambda}^{\hat{\alpha}}\hat{\omega}_{\hat{\beta}}S^{\beta\hat{\beta}}_{\alpha\hat{\alpha}}+S_{gh}\big]\,.\nonumber
\end{eqnarray}
In this way we have split the action into four sectors depending of their ghost content.
We are almost ready to read the target space superfields by comparing the  action described above with the deformed action \eqref{deformado} 
and to show that the deformation  of the PS $AdS_{5}\times S^{5}$ superstring yields the same target space supergeometry as the homogeneous YB deformation of the GS $AdS_{5}\times S^{5}$ superstring \cite{target}.

The YB deformations of the $GS$ superstring  \cite{vicedo,kawa} is implemented through the Lie algebra operator
 \begin{eqnarray}
 \mathcal{O}_{GS-}=1-\eta R_{g}\circ d_{GS}\,,\quad \mathcal{O}_{GS+}=1+\eta R_{g}\circ \hat{d}_{GS}\,.
 \end{eqnarray}
Their components are linear combinations of the projectors
 \begin{equation}
 d_{GS}=P_{1}+2\hat{\eta}^{2}P_{2}-P_{3}\,,\quad \hat{d}_{GS}=-P_{1}+2\hat{\eta}^{2}P_{2}+P_{3}\,,
 \end{equation}
where $\hat{\eta}=(1-c\eta^{2})^{1/2}$. Since we are interested in the case when $R$ satisfies the CYBE this means that $\hat{\eta}=1$.
It is convenient to define the GS deformed currents
\begin{equation}
J_{GS\pm}=\mathcal{O}_{GS\pm}A\,,
\end{equation}
so that the action of the GS $\eta$-model is written as
\begin{equation}
S_{GS}=-\frac{1}{4}(\gamma^{ij}-\epsilon^{ij})\int Str(A_{i},d_{GS}J_{GS,j})\,.
 \label{eta}
 \end{equation}
A nice approach was introduced in \cite{target} in order to read the target space supergeometry. In particular, the supervielbiens $E^{A}$ of the deformed geometry are given by
 \begin{equation}
 E_{2}^{a}=J_{GS2+}^{a}\,,\quad E_{1}^{\alpha}=Ad_{h}J_{GS1+}^{\alpha}\,,\quad E_{3}^{\hat{\alpha}}=J_{GS3-}^{\hat{\alpha}}\,,
 \label{vielbein}
 \end{equation}
 where $h$ is an element of the isotropy group $SO(5,1)\times SO(6)$.
 
To illustrate the correspondence between the two superstrings it is worthwhile to rewrite \eqref{deformado} in GS language. 
For this purpose we define the operators
\begin{equation}
\vartheta_{\pm}=\mathcal{O}^{-1}_{PS\pm}\mathcal{O}_{GS\pm}\,,
\label{teta}
\end{equation}
which relates the PS deformed currents $J_{\pm}$ defined in \eqref{416} to $J_{GS}$ as
\begin{equation}
J_{\pm}=\vartheta_{\pm}J_{GS\pm}\,,
\end{equation}
It is useful to keep in mind some useful identities involving  \eqref{teta}:
\begin{eqnarray}
\vartheta_{-}=1-\frac{4}{3}P_{3}+\frac{4}{3}\mathcal{O}^{-1}_{PS-}P_{3}\,,\qquad
\vartheta^{-1}_{-}=1-4P_{3}+4\mathcal{O}^{-1}_{GS-}P_{3}\,,\nonumber\\
\vartheta_{+}=1-\frac{4}{3}P_{1}+\frac{4}{3}\mathcal{O}^{-1}_{PS+}P_{1}\,,
\qquad
\vartheta^{-1}_{+}=1-4P_{1}+4\mathcal{O}^{-1}_{GS+}P_{1}\,.
\label{iden}
\end{eqnarray}
We start by examining the matter sector of \eqref{deformado}. This contribution should be compared with the matter sector coming from \eqref{integrada} which will allow us to read the background fields $B$ and $P_{\alpha\hat{\alpha}}$. It reads
\begin{eqnarray}
&\frac{1}{4}&Str(A_{+},d_{PS}J_{-})\\
&=&\frac{1}{4}Str(\bar{J}_{GS-},d_{PS}\circ (1-\frac{4}{3}P_{3})J_{GS-})
+\frac{1}{4}Str(\bar{J}_{GS-},\hat{d}_{GS}\circ R_{g}\circ d_{PS} (1-\frac{4}{3}P_{3})J_{GS-})+\nonumber\\
&+&\frac{4}{3}Str(\bar{J}_{GS-},d_{PS}\circ\mathcal{O}^{-1}_{PS-}P_{3}J_{GS-})+\frac{4}{3}Str(\bar{J}_{GS-},(\eta\hat{d}_{GS}\circ R_{g})\circ d_{PS}\circ\mathcal{O}^{-1}_{PS-}P_{3}J_{GS-})\,.\nonumber
\end{eqnarray}
Notice that $d_{PS}\circ (1-\frac{4}{3}P_{3})=d_{GS}$. After a convenient rearrangement of the last term, using $J_{GS-}=(\mathcal{O}^{-1}_{GS-}\circ\mathcal{O}_{GS+})J_{GS+}$, the matter part of the action takes the following form
\begin{eqnarray}
\frac{1}{4}Str(\bar{A},d_{PS}J_{PS-})&=&\frac{1}{4}Str(\bar{J}_{GS-},d_{GS}J_{GS-})
+\frac{1}{4}Str(\bar{J}_{GS-},\hat{d}_{GS}\circ R_{g}\circ d_{GS} J_{GS-})+\nonumber\\
&+&\frac{4}{3}Str(P_{3}J_{GS-},P_{1}\circ \vartheta_{+}P_{1}\bar{J}_{GS+})\,,
\label{gs}
\end{eqnarray}
 The first two terms of \eqref{gs} can be compared with the first two terms of \eqref{integrada}. so that the metric and the $B$-field are
\begin{equation}
G_{MN}\partial Z^{M}\bar{\partial}Z^{N} = Str(\bar{J}_{GS-},J_{GS-})\,,\quad B=\frac{1}{2}(P_{1}-P_{3}+\eta\hat{d}_{GS}\circ R_{g}\circ d_{GS})\,.
\end{equation}
They match the metric and $B$ field of the GS YB deformations \cite{target}.

Now we look for the Ramond-Ramond bispinor. Substituting \eqref{vielbein} in the last term in \eqref{gs}, we have 
\begin{eqnarray}
\frac{2}{3}Str(E_{3}^{\hat{\alpha}}t^{3}_{\hat{\alpha}},P_{1}\circ \vartheta_{+}P_{1}\circ Ad^{-1}h\bar{E}_{1}^{\alpha}t^{1}_{\alpha})\,,
\end{eqnarray}
which can be compared with \eqref{integrada} to get
\begin{equation}
P_{\alpha\hat{\alpha}}=\frac{1}{2} (P_{1}\circ \vartheta_{+}P_{1}\circ Ad^{-1}_{h})_{\alpha}^{\beta}\mathcal{K}_{\hat{\alpha}\beta}\,,
\end{equation}
where $\mathcal{K}_{\hat{\alpha}\alpha}=Str(t^{3}_{\hat{\alpha}},t^{1}_{\alpha})$. Taking into account \eqref{iden} we can write $P^{\alpha\hat{\alpha}}$ as
\begin{eqnarray}
P^{\alpha\hat{\alpha}}= 2\mathcal{K}^{\hat{\alpha}\beta} (Ad_{h}\circ \vartheta^{-1}_{+} )_{\alpha}^{\beta}=8\mathcal{K}^{\hat{\alpha}\beta} (Ad_{h}\circ (3-4\mathcal{O}^{-1}_{GS+}) )^{\alpha}_{\beta}\,,
\end{eqnarray}
which is the same RR bispinor found for the homogeneous YB deformations  \cite{target}. Now we move to the matter-ghost sector.  First of all we notice that  
\begin{eqnarray}
J^{0}_{PS-}&=&P_{0}(1-\frac{4}{3}P_{3}+\frac{4}{3}\mathcal{O}^{-1}_{PS-}\circ P_{3}) J_{GS-}=J^{0}_{GS-}+\frac{4}{3}P_{0}\circ\mathcal{O}^{-1}_{PS-} J^{3}_{GS-}\,,\nonumber\\
J^{0}_{PS+}&=&P_{0}(1-\frac{4}{3}P_{1}+\frac{4}{3}\mathcal{O}^{-1}_{PS+}\circ P_{1}) J_{GS-}=J^{0}_{GS+}+\frac{4}{3}P_{0}\circ\mathcal{O}^{-1}_{PS+} J^{1}_{GS+}\,.
\end{eqnarray}
Using these equations in the matter-ghost sector of \eqref{deformado} we have 
\begin{eqnarray}
Str(J_{0-},N_{0+})+Str(\bar{J}_{0+},N_{0-})&=&
Str(J^{0}_{GS-},N_{0+})+\frac{4}{3} Str(P_{0}\circ\mathcal{O}^{-1}_{PS-} J^{3}_{GS-},N_{0+})\nonumber\\&+&Str(\bar{J}^{0}_{GS+},N_{0-})+\frac{4}{3} Str(P_{0}\circ\mathcal{O}^{-1}_{PS+} \bar{J}^{1}_{GS+},N_{0-})\,.\nonumber\\
\end{eqnarray}
The above equation can be compared with the second line of \eqref{integrada} to read the spin connection $(\Omega,\hat{\Omega})$ and the pair $(C, \tilde{C})$ 
\begin{eqnarray}
\Omega^{ab}=J_{GS-}^{ab}\,,&\qquad&
C^{\beta\hat{\gamma}}_{\sigma}= 4 (Ad_{h}\circ (3-4\mathcal{O}^{-1}_{GS+}) )^{\gamma}_{\beta}\mathcal{K}^{\hat{\alpha}\beta}(\mathcal{O}^{-1}_{PS-})_{\alpha}^{ab} (\gamma_{ab})_{\sigma}^{\beta}\,,
\\
\hat{\Omega}^{ab}=\bar{J}_{GS+}^{ab}\,,&\qquad&
\tilde{C}^{\hat{\beta}\gamma}_{\hat{\sigma}}=4 (Ad_{h}\circ (3-4\mathcal{O}^{-1}_{GS+}) )^{\gamma}_{\beta}\mathcal{K}^{\hat{\alpha}\beta} (\mathcal{O}^{-1}_{PS-})_{\hat{\alpha}}^{ab} (\gamma_{ab})_{\hat{\sigma}}^{\hat{\beta}}\,.
\end{eqnarray}
Finally, the same analysis can be done for the $N_{0-}N_{0+}$ sector and we find 
\begin{equation}
S^{\alpha\hat{\alpha}}_{\beta\hat{\beta}}=(\gamma_{ab})_{\hat{\beta}}^{\hat{\alpha}}(\gamma^{cd})^{\alpha}_{\beta}(1-\eta \mathcal{O}^{-1}_{PS-}R_{g})^{ab}_{cd}\,.
\end{equation}
This shows that the homogeneous YB deformations of the GS superstring  and our deformation of pure spinor in $AdS_{5}\times S^{5}$ have the same geometry and target space contents, that is, the same generalized supergravity background. 

\section{Concluding remarks}

We have shown how to build  homogeneous YB deformations for the PS superstring in $AdS_{5}\times S^{5}$ by exploiting its BRST symmetry and using homological perturbation theory. Even though we restricted our analysis to the case where the $R$-matrix satisfies the homogeneous CYBE, the extension to the non-homogeneous case should proceed along the same lines as in \cite{Bevilaqua} and we expect a simple relation between them as in the case of deformed GS superstrings \cite{vicedo,kawa}.

We have found a one to one correspondence between deformations of the action and the cohomology of the BRST charge for the PS superstring in $AdS_{5}\times S^{5}$. Having found the deformed BRST operator it would be interesting to study the elements in its cohomology as, for instance, the deformation of the dilaton vertex operator considered in \cite{Berkovits:2008ga}.

It is important to remark that our analysis is completely classical and it is plausible to expect that extra quantum requirements may enforce on-shell supergravity. As it was shown in \cite{scale} the YB deformed GS model preserves the original scale invariance and defines a UV finite theory. In the PS case we expect that the Weyl symmetry can only be recovered when the deformed target space allows a type IIB supergravity solution suggesting that the central charge of the deformed model must be proportional to the unimodular condition for the $R$-matrix.

As remarked in the introduction, the GS superstring propagates in a background which is restricted by  $\kappa$-symmetry to be a solution of generalized supergravity  \cite{kappa}. Our results strongly suggest that, at least classically, the constraints  imposed on the target superspace by the nilpotency and holomorphicity of the BRST charge \cite{BH} should also imply the equations of motion for generalized supergravity. As in the case of the GS superstring \cite{scale}, we expect that this condition would be, in principle, sufficient to get a vanishing one-loop beta function.
 
\acknowledgments
We would like to thank Nathan Berkovits and Andrei Mikhailov for useful comments and suggestions.
The work of H.A.B. was supported by CAPES and the work of V.O.R. was supported by FAPESP grants 2014/18634-9 and 2016/01343-7.

\appendix 
\section{Some properties of $\mathcal{P}$}
First of all we note that
\begin{eqnarray}
Str(\mathcal{P}_{13}A_{1},A_{3})=Str(A_{1},\mathcal{P}_{31}A_{3})\,.
\end{eqnarray}
In order to prove some important properties of $\mathcal{P}$ we use the following theorem (see section 7 in \cite{Bevilaqua})
\begin{theorem}
If $[\lambda_{1},[\lambda_{3},S_{2}]]=0$, for any $S_{2}$, then it implies that $[\lambda_{3},S_{2}]=0$. Analogously, if $[\lambda_{3},[\lambda_{1},S_{2}]]=0$, then $[\lambda_{1},S_{2}]=0$.\end{theorem}

\begin{prop} $Q_{0L}\mathcal{P}_{13}(gt_{a}g^{-1})_{1}=0$\\
\textit{Proof.}
\begin{eqnarray}
0=Q_{0L}[\lambda_{1},\mathcal{P}_{13}(gt_{a}g^{-1})]&=&[\lambda_{1},Q_{0L}\mathcal{P}_{13}(gt_{a}g^{-1})]\,,\nonumber\\
&=&[\lambda_{1},[\lambda_{3},(gt_{a}g^{-1})_{1}+Q_{0L}S_{2}]]
\end{eqnarray}
Hence, from the above theorem
\begin{eqnarray}
0=[\lambda_{3},(gt_{a}g^{-1})_{1}+Q_{0L}S_{2}]&=&Q_{0L}((gt_{a}g^{-1})_{1}+[\lambda_{3},S_{2}])\,,
\end{eqnarray}
Hence, $Q_{0L}\mathcal{P}_{13}(gt_{a}g^{-1})_{1}=0$.
\end{prop}
\begin{prop}$Q_{0R}\mathcal{P}_{13}(gt_{a}g^{-1})_{1}=[\lambda_{1},(gt_{a}g^{-1})_{0}]$\\
\textit{Proof.}
\begin{eqnarray}
0=Q_{0R}[\lambda_{1},\mathcal{P}_{13}(gt_{a}g^{-1})_{1}]&=&[\lambda_{1},[\lambda_{1},(gt_{a}g^{-1})_{1}]+[\lambda_{3},Q_{0R}S_{2}]]\,,
\end{eqnarray}
As shown above, this equality implies that
\begin{equation}
0=[\lambda_{1},Q_{0R}[\lambda_{3},S_{2}]]\, \implies Q_{0R}[\lambda_{3},S_{2}]=0,\\
\end{equation}
and it follows that $Q_{0R}\mathcal{P}_{13}(gt_{a}g^{-1})_{1}=[\lambda_{1},(gt_{a}g^{-1})_{0}]$.
\end{prop}
Analogously, it can be proved that
\begin{prop}
\begin{equation}
Q_{0R}\mathcal{P}_{31}(gt_{a}g^{-1})_{3}=0\,.
\end{equation}
\end{prop}

\begin{prop}
\begin{equation}
Q_{0L}\mathcal{P}_{31}(gt_{a}g^{-1})_{3}=[\lambda_{3},(gt_{a}g^{-1})_{0}]\,.
\end{equation}
\end{prop}



\bibliographystyle{unsrt}



\end{document}